\documentclass[12pt]{article}
\usepackage{rotate}
\usepackage{epsfig}
\usepackage{psfrag}
\usepackage{a4}

\begin{document}

\newcommand \be  {\begin{equation}}
\newcommand \bea {\begin{eqnarray} \nonumber }
\newcommand \ee  {\end{equation}}
\newcommand \eea {\end{eqnarray}}

\title{\bf More statistical properties of order books and price impact}
\author{Marc Potters$^*$, Jean-Philippe Bouchaud$^{\dagger,*}$}
\maketitle
{\small
{$^\dagger$ Commissariat \`a l'Energie Atomique, Orme des Merisiers}\\
{91191 Gif-sur-Yvette {\sc cedex}, France}\\

{$^*$ Science \& Finance, CFM, 109-111 rue Victor Hugo}\\
{92 353 Levallois {\sc cedex}, France}\\

\date{\today}
}

\begin{abstract}
We investigate present some new statistical properties of order books. 
We analyse data from the Nasdaq and investigate (a) the statistics of 
incoming limit order prices, (b) the shape of the average order book, and (c)
the typical life time of a limit order as a function of the distance 
from the best price. We also determine the `price impact' function using 
French and British stocks, and find a logarithmic, rather than a power-law,
dependence of the price response on the volume. The weak time 
dependence of the response function shows that the impact is, surprisingly, 
quasi-permanent, and suggests that trading itself is interpreted by the
market as new information. 
\end{abstract}


Many statistical properties of financial markets have already been explored, 
and have
revealed striking similarities between very different markets (different traded assets,
different geographical zones, different epochs) \cite{Cont,MS,BP}. More
recently, the statistics of the `order book', which is the ultimate 
`microscopic' level of description of financial
markets, has attracted considerable attention, both from an empirical 
\cite{Biais,Maslov,Challet,us,Farmer1} and theoretical 
\cite{Bak,Kogan,Maslov,Aussie,Slanina,Farmer2,us,Challet2,Challet3} point of view.    

The order book is the list of all buy and sell limit orders, with their 
corresponding price
and volume, at a given instant of time. We will call $a(t)$ the ask price (best sell
price) at time $t$ and $b(t)$ the bid price (best buy price) at time $t$. 
The midpoint $m(t)$ is the average between the bid and the ask: $m(t)=[a(t)+b(t)]/2$. 
When a new order appears 
(say a buy order), it either adds to the book 
if it
is below the ask price, or generates a trade at the ask if it is above 
(or equal to) the ask 
price (we call these `market orders').
The price dynamics is therefore the result of the interplay 
between the order book and the order flow. The study of the order book is very 
interesting both for academic and practical reasons. It provides intimate 
information on the processes of trading and price formation, and reveals a 
non trivial structure of the agents expectations: as such, it is 
of importance to test some basic notions of economics. The practical 
motivations are also obvious: issues such as the market impact or 
the relative merit of 
limit versus market orders are determined by the structure and dynamics of the
order book.

The main results of our investigation of some major French stocks were as follows 
\cite{us}: 
(a) the price at which new limit orders are placed is,
somewhat surprisingly, very broadly (power-law) distributed around the current bid/ask; 
(b) the average order book has a maximum away from the current bid/ask, and a tail 
reflecting the
statistics of the incoming orders. We studied numerically
a `zero intelligence' model of order book which reproduces most of the 
empirical results, and proposed a simple approximation to compute analytically 
the characteristic humped shape of the average order book (see also \cite{Farmer2}). 

In this paper, we give the results concerning some of the Nasdaq order books, 
as observed on the Island ECN (see www.island.com), and discuss the similarities
and differences with the French data. Second, we give some results on the 
price impact function that quantifies how a transaction of a given volume 
affects the price (on average).   

\section{Results on Nasdaq stocks}

We denote by $b(t)-\Delta$ the
price of a new buy limit order, and $a(t)+\Delta$ the price of a new sell limit order. 
A first interesting question concerns the distribution 
density of $\Delta$, i.e. the distance 
between the current price and the incoming limit order. We found that $P(\Delta)$ 
for French stocks was identical
for buy and sell orders (up to statistical fluctuations); and very well fitted 
by a single power-law:
\begin{equation}
\label{PDelta}
P(\Delta) \propto \frac{\Delta_0^\mu}{(1+\Delta)^{1+\mu}},
\end{equation}
with an exponent $\mu \simeq 0.6$. This power-law was confirmed in \cite{Farmer1}
for British stocks, albeit with a different exponent $\mu \simeq 1.5$. Note however
that all the volume is electronic in Paris, which is not the case in London.

When repeating this analysis for some of the Nasdaq stocks, we found results 
that significantly depend on the studied asset (see Fig. 1), but that all reveal 
the very slowly 
decaying tail discovered in the case of French stocks. As emphasized in 
\cite{us,Farmer1}, this tail suggests that market participants believe 
that large jumps in the 
price of stocks 
are always possible, and place orders very far from the current price in order to
take advantage of these large potential fluctuations. 

\begin{figure}
\begin{center}
\epsfig{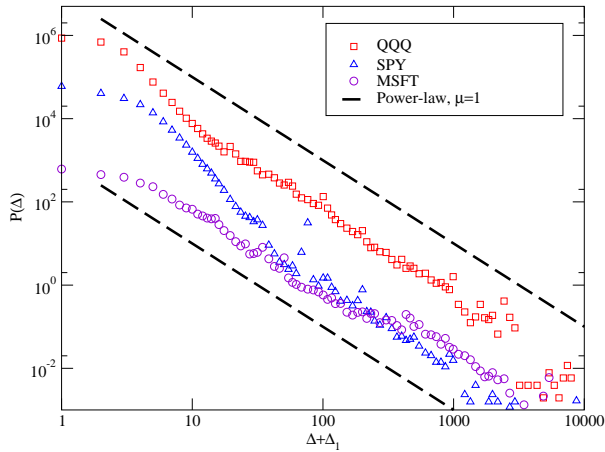}
\end{center}
\caption{\small Cumulative distribution of the position $\Delta$ of incoming
orders, as a function of $1+\Delta$ (in ticks), for QQQ, SPY and MSFT. The
dashed lines correspond to $\mu=1$. \label{fig1} }
\end{figure}

More precisely, we have studied QQQ and SPY, which are exchange traded funds that
track, respectively, the Nasdaq and the S\&P500, and MSFT (Microsoft). The 
data corresponds to the period June 1st to July 15th, 2002. 
At variance with the Paris Bourse, where all the volume is traded on a
centralized electronic market, the Nasdaq is in fact a myriad of electronic
platforms. Island is one of them, which only gathers a fraction of the total
volume (roughly 40 \% for QQQ and 20 \% for SPY and MSFT). Therefore, unfortunately,
the statistics that we report here only contains a partial information on the
order flow and on the order book. This is why this data is less representative than
that on French stocks. In the case of QQQ however, the Island ECN is considered to
be the `dominant' market, which drives all other platforms; the 
data reported below is therefore probably significative in this case. 
Fig. 1 shows that for all three assets, the tail index $\mu$ of 
$P(\Delta)$ is in the same ballpark as the values found for the French and British 
stocks. One can notice that the value of $\mu$ for MSFT is smaller than that for 
QQQ, itself smaller than gor SPY. This is expected, since large jumps are 
more probable for individual stocks than it is for the Nasdaq index, itself more 
volatile than the S\&P500.

We now turn to the shape of the order book. The order flow is maximum around the
current price, but 
an order very near to the current price has a larger probability to be executed or
cancelled (see below) and disappear from the book. 
It is thus not a priori clear what will be the 
shape of the average order book. We find that in the case of QQQ the (time-averaged)
size of the queue in order book is symmetrical, and has a maximum away from the 
current bid (ask), as was found for French stocks: see Fig. 2. For SPY (and for 
MSFT, not shown), on the other hand, the size of the queue is maximum at the bid 
(or ask). The difference might be due to the fact that, as explained above, the
Island ECN is not the dominant player for SPY or MSFT, and often behaves as a 
`mirror' of other platforms.   

\begin{figure}
\begin{center}
\epsfig{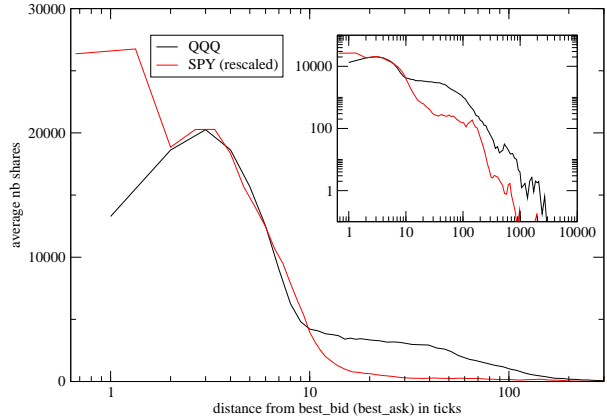}
\end{center}
\caption{\small Average order book for QQQ and SPY, as a function of the
distance $\Delta$ from the current bid (or ask). The axis have been rescaled in the
case of SPY.  \label{fig2} }
\end{figure}

In our previous paper \cite{us}, we have analysed a simple model that explains 
the humped shape
of the book of the type observed on QQQ. The basic ingredients of this model is (a) the
non uniform, power-law like, flow of incoming orders (b) the diffusive like dynamics
of the price that eats up the nearby limit orders and (c) the finite life-time of 
the limit orders. For simplicity, this life-time was assumed in \cite{us} to be
constant, independent of the distance form the best price $\Delta$. We have 
investigated empirically this question using the Nasdaq data (the French data 
unfortunately does not provide the time at which a limit order is cancelled).
As a proxy for the cancel rate, we have computed the number of shares 
cancelled per unit time as a function of $\Delta$, and divided the result by the 
average number of shares in the order book at distance $\Delta$. The results for 
QQQ and MSFT are shown in Fig. 3. One sees that the life-time of a given order
increases as one moves away from the bid-ask. This is, again, expected from the 
arguments given in \cite{us,Farmer1}: far 
away orders are typically put in the market by patient investors that want to
take profit of important swings in the medium term. Orders at and around the bid
and ask prices, on the other hand, correspond to very active market participants 
that observe the market price all the time and readjust their orders at a very 
high frequency. Our results suggest that any quantitative theory of the order 
book should include this non uniform cancel rate.  

\begin{figure}
\begin{center}
\epsfig{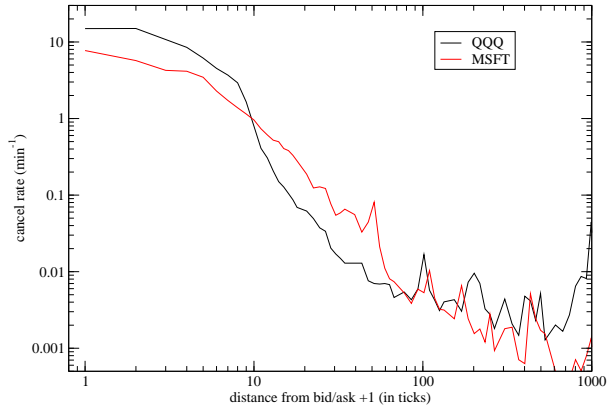}
\end{center}
\caption{\small Cancel rate for QQQ and MSFT, as a function of the
distance $\Delta$ from the current bid (or ask). Note that the cancel rate at the 
bid/ask is very high (10 per minute), which suggests that most of the orders are 
automated. Note that the execution rate is only $22 \%$ of the cancel rate for QQQ, and
$40 \%$ for MSFT. \label{fig3} }
\end{figure}

\section{The price impact function}

Recently, several studies have tried to determine quantitatively how a market order
of a given volume affects the price. This information is extremely important for
many purposes. First, for model building: many agent based models of markets 
use as a starting point a phenomenological relation between price changes and 
order imbalance \cite{Beja,Farmer3,BC,Gabaix}. Second, as far as trading 
is concerned, the
control of market impact is crucial whan one wants to manage large volumes. 

The most naive idea, inspired from physical systems, is that of linear 
response: prices should move proportionally to volumes. However, some recent 
work \cite{Germans,Gopi2,Farmer4} show that the average price change 
$\Delta p$ is a sublinear function of the 
volume imbalance $\Delta V$. A square-root dependence 
$\Delta p \propto \Delta V^\alpha$ with $\alpha=1/2$ was advocated on 
theoretical grounds \cite{Zhang,Gabaix}. An exhaustive study of all US stocks 
\cite{Farmer4} however
suggests a smaller exponent $\alpha$ but a log-log plot of $\Delta p$ vs. $\Delta V$ 
reveals a systematic downward bend, which indicates that a power-law might not be 
appropriate. Here, we want to argue that this relation might in fact be logarithmic.

More precisely, let us define the time dependent response function $R(V,\tau)$ 
as the average mid-point variation between times $t$ and $t+\tau$, {\it 
conditioned} to a transaction of volume $V$ taking place at time $t$ at the 
ask (buyer initiated trade, $\varepsilon(t)=+1$) or at the bid (seller initiated trade, 
$\varepsilon(t)=-1$). The response function is given by:
\be
R(V,\tau)= \left\langle \varepsilon(t)\cdot \left[m(t+\tau)-m(t)\right] | V \right \rangle.
\ee
[We have actually studied the full distribution of 
$\varepsilon(t) \left[m(t+\tau)-m(t)\right]$ and checked that the 
above average is 
{\it not} dominated by a few market `jumps'.]
The results we find (see Fig. 4) can be expressed as:
\be
R(V,\tau) \approx {\cal R}(\tau) \ln V,\label{log}
\ee 
where, surprisingly, ${\cal R}(\tau)$ is only weakly dependent on $\tau$: 
${\cal R}(\tau)$ first increases from $\tau=10$ seconds to a few hundred seconds, 
and then appears to decrease back to a finite value, with a total amplitude of 
variation of at most $50\%$. This means that (i) as found in previous
investigations, the impact of small trades on the price is, 
in relative terms, much larger (statistically)  than that of large trades, 
(ii) that the impact 
of trading on the price is quasi-permanent, and (iii) the (weak) temporal structure of 
${\cal R}(\tau)$ is compatible with the observed price dynamics, which is found to be
(weakly) super-diffusive at very short times and (weakly) sub-diffusive on 
intermediate time scales.

\begin{figure}
\begin{center}
\epsfig{file=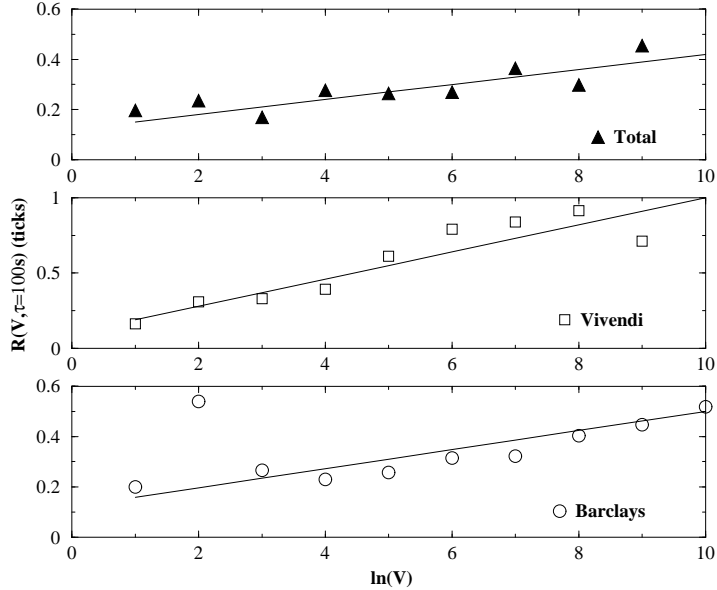,width=8cm,angle=270}
\end{center}
\caption{\small Price response function $R(V,\tau)$ as a function of $\ln V$ 
for three stocks (Barclays, Vivendi and Total). The period corresponds to February 
2002 for the French stocks and to May-June 2002 for Barclays. The time delay 
$\tau$ is equal to $100$ seconds. The $y$-axis is in ticks.   \label{fig4} }
\end{figure}

Point (i) is perhaps not as surprising as it first seems: as emphasized above, the
average queue in the order book is an increasing function of $\Delta$. This means 
that price changes experience a stronger resistance from the book for large volumes
than for small volumes. In other words, since the most probable volume at the 
bid (or at the ask) is one share, a small transaction is capable of changing the
mid-price by eating up the bid (or ask). The next tick (bid minus one or ask plus one)
has typically a larger volume waiting, and offers therefore more resistance to
further price changes. The humped
shape of the order book in itself is, as recently emphasized in \cite{Farmer4}, enough
to give a downward bend to $R(V,\tau)$ vs. $V$. However, this effect is not 
sufficient to account for the very slow $\ln V$ behaviour reported above. The
interpretation is that, most probably, market orders of large volumes are only 
submitted when the order book on the opposite side of the trade is capable of 
absorbing this
incoming volume. We have indeed found some positive correlation between the volume at 
bid/ask before the trade and the volume of the following trade.

Point (ii) is in fact quite intriguing, in particular in the context of the efficient 
market hypothesis. One might have expected that, in the absence 
of any new information, there should be a restoring force driving back the new
mid-point towards a local `equilibrium' price. The near-absence of temporal structure 
in ${\cal R}(\tau)$ suggests that each new trade is in fact interpreted by the 
market {\it as new information}, and the new mid-point is immediately adopted
as the new reference price, around which the flow of incoming orders readapts. 
(This is actually the hypothesis made in the models investigated in \cite{Farmer2,us}) 
This reverts the usual logic: most of the time, prices move not because of new 
information, but rather new information is generated by the mere changes of prices.
Although there seems to be a small resilience on intermediate time
scales, the prices are much more affected by trading itself than expected from 
the theory of efficient market. This observation might explain the tremendous 
`excess' volatility of financial prices.

\vskip 1cm
\noindent{Acknowledgements:} We thank Jean-Pierre Aguilar, Jelle Boersma, 
Damien Challet, J. Doyne Farmer, Xavier Gabaix, Andrew Matacz, Rosario Mantegna, 
Marc M\'ezard and Matthieu Wyart for stimulating and useful discussions.


\end{document}